\documentstyle[supertabular,epsf,rotate]{mn}
\newcommand{\mc}{\multicolumn}

\newcommand{\ltsimeq}{\raisebox{-0.6ex}{$\,\stackrel 
        {\raisebox{-.2ex}{$\textstyle <$}}{\sim}\,$}} 
\newcommand{\gtsimeq}{\raisebox{-0.6ex}{$\,\stackrel 
        {\raisebox{-.2ex}{$\textstyle >$}}{\sim}\,$}} 

\begin{document}

\title[The quasar fraction in low--frequency selected complete
samples] {The quasar fraction in low--frequency selected complete
samples and implications for unified schemes}

\author[Willott et al.]{Chris J.\ Willott$^{1,2}$\footnotemark, Steve Rawlings$^{1}$,
Katherine M.\ Blundell$^{1}$ and Mark Lacy$^{1,3}$\\ 
$^{1}$ Astrophysics, Department of Physics, Keble Road, Oxford, OX1
3RH, U.K. \\
$^{2}$ Instituto de Astrof\'\i sica de Canarias, C/ Via Lactea s/n, 38200
La Laguna, Tenerife, Spain \\
$^{3}$ IGPP, L-413, Lawrence Livermore National Laboratory, Livermore, CA 94550, 
USA \\}

\maketitle

\begin{abstract}

Low-frequency radio surveys are ideal for selecting
orientation-independent samples of extragalactic sources because the
sample members are selected by virtue of their isotropic
steep-spectrum extended emission. We use the new 7C Redshift Survey
along with the brighter 3CRR and 6C samples to investigate the
fraction of objects with observed broad emission lines -- the `quasar
fraction' -- as a function of redshift and of radio and narrow
emission line luminosity. We find that the quasar fraction is more
strongly dependent upon luminosity (both narrow line and radio) than
it is on redshift. Above a narrow [OII] emission line luminosity of
$\log_{10}(L_{\mathrm [OII]}/{\mathrm W}) \gtsimeq 35$ [or radio
luminosity $\log_{10} (L_{151} /$ W Hz$^{-1}$ sr$^{-1}) \gtsimeq
26.5$], the quasar fraction is virtually independent of redshift and
luminosity; this is consistent with a simple unified scheme with an
obscuring torus with a half-opening angle $\theta_{\rm trans} \approx
53^{\circ}$. For objects with less luminous narrow lines, the quasar
fraction is lower. We show that this is not due to the difficulty of
detecting lower-luminosity broad emission lines in a less luminous,
but otherwise similar, quasar population.  We discuss evidence which
supports at least two probable physical causes for the drop in quasar
fraction at low luminosity: (i) a gradual decrease in $\theta_{\rm
trans}$ and/or a gradual increase in the fraction of lightly-reddened
($0 \ltsimeq A_{V} \ltsimeq 5$) lines-of-sight with decreasing quasar
luminosity; and (ii) the emergence of a distinct second population of
low luminosity radio sources which, like M87, lack a well-fed quasar
nucleus and may well lack a thick obscuring torus.

\end{abstract}

\begin{keywords}
galaxies:$\>$active -- galaxies:$\>$nuclei -- quasars:$\>$general 
-- galaxies:$\>$evolution 
\end{keywords}

\footnotetext{Email: cjw@astro.ox.ac.uk}

\section{Introduction}

Orientation-based unified schemes for radio-loud quasars and powerful
radio galaxies play a key role in our understanding of these
objects. Ever since the conception of the idea that powerful radio galaxies are
simply quasars with their jet axes oriented away from our line of
sight, such that the nuclear continuum and broad-line regions are
obscured by a dusty torus or warped disc (Scheuer 1987; Barthel 1989),
evidence has been sought to prove that these unification schemes are
correct. Reviews of unification schemes for active galactic nuclei
have been presented by Antonucci (1993) and Urry \& Padovani
(1995). Although the majority of observations are consistent with
them, some observations have been used to cast doubt on their 
viability over all ranges of radio luminosities and redshifts.

The narrow line emission in radio sources is observed to be emitted
largely from beyond the obscuring material (e.g. McCarthy, Spinrad \&
van Breugel 1995; Hes, Barthel \& Fosbury 1996) and therefore is
independent of the jet axis orientation. Hence in this model one would
expect radio galaxies and quasars to have similar narrow line
luminosities. The strong positive correlation between the extended
radio luminosities and narrow line luminosities of radio sources (Baum
\& Heckman 1989; Rawlings et al. 1989; Willott et al. 1999) means that
the samples of radio galaxies and quasars to be compared must be
matched in extended radio luminosity. At low redshift ($z<0.8$), there
have been claims that quasars are observed to have [OIII] line
luminosities a factor of 5-10 greater than radio galaxies of similar
radio luminosities (Baum \& Heckman 1989; Jackson \& Browne 1990;
Lawrence 1991), although the [OII] luminosities of radio galaxies and
quasars at these redshifts are indistinguishable (Browne \& Jackson
1992; Hes et al. 1996). These differences between [OII] and [OIII]
have been interpreted as being due to partial obscuration of [OIII]
since its higher ionization potential means it is likely to be emitted
from closer to the nucleus than [OII]. However, Jackson \& Rawlings
(1997) have investigated the [OIII] luminosities of $z>1$ radio
galaxies and quasars and find their distributions
indistinguishable. Using a combined 7C/3CRR dataset Willott et
al. (1999) find that quasars have more luminous narrow lines than
radio galaxies at intermediate radio luminosities [$26<\log_{10}
(L_{151}$ / W~Hz$^{-1}$sr$^{-1}) <27$], but they are similar at higher
radio luminosities; this result leads to different slopes for the
narrow-line versus radio luminosity correlation for quasars and radio
galaxies.

Barthel (1989) showed that in the redshift range $0.5<z<1.0$ the
linear size distributions and fraction of quasars in the 3CRR sample
are consistent with all radio galaxies having their jet axes at an
angle $\theta_{\rm trans}>44^{\circ}$ from our line-of-sight and all
the quasars having $\theta_{\rm trans}<44^{\circ}$, where $\theta_{\rm
trans}$ is the half-opening angle of the obscuring torus; a value of
$\theta_{\rm trans}=45^{\circ}$ has been adopted by many on the basis
of this paper.  However, at higher redshifts in the 3CRR sample, the
fraction of quasars increases giving $\theta_{\rm trans} \approx
60^{\circ}$ (e.g. Singal 1993). Note that because of the tight
luminosity--redshift correlation inherent in a single flux-limited
sample, this apparent correlation with redshift may be due instead to
a correlation with radio luminosity. Singal (1996) used several large
samples with differing radio flux-density limits to find that the
quasar fraction in radio samples declines with decreasing radio
flux-densities. However, due to the correlation between the optical
and radio luminosities of steep-spectrum quasars (Serjeant et
al. 1998; Willott et al. 1998a), radio-fainter samples will contain
optically-fainter quasars. Therefore the quasar fraction in faint
samples might be underestimated if only quasars brighter than a
certain optical magnitude limit are identified, which may well be the
case in some of these fainter samples used by Singal.

Using the new 7C Redshift Survey, a low-frequency radio sample
selected at a flux-density limit $25 \times$ lower than the 3CRR sample with
$ \approx 90\%$ spectroscopic redshift completeness, we have
previously shown that the quasar fraction does not depend strongly
upon radio luminosity or redshift for $1<z<3$ (Willott et
al. 1998b). Note that the low-frequency selection of the samples is
crucial, because this ensures selection on {\em extended} radio flux
which should be entirely isotropic. In this paper we use a larger
sample to investigate how the quasar fraction of radio sources depends
upon redshift, radio luminosity and narrow emission line luminosity
and we consider the implications for orientation-based unified schemes. In a
companion paper (Blundell, Rawlings \& Willott in prep.) we will
investigate the constraints placed on unified schemes by
the radio properties of quasars and radio galaxies 
from the combined 7C/6C/3CRR dataset. We assume throughout this paper
that $H_{\circ}=50~ {\rm km~s^{-1}Mpc^{-1}}$ and $q_{\circ}=0.5$.

\section{The sample}

\begin{figure*}
\hspace{0.8cm} \epsfxsize=0.93\textwidth \epsfbox{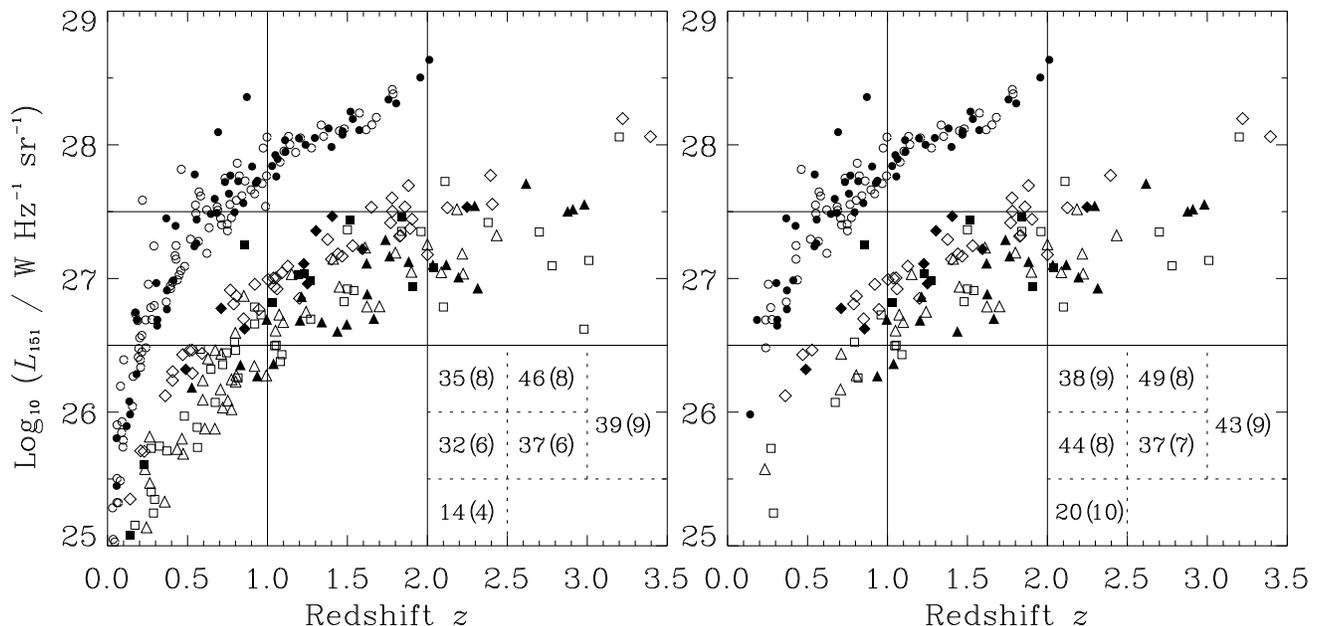}
{\caption[junk]{\label{fig:qf1}The radio luminosity--redshift plane
for the 3CRR (circles), 6C (diamonds), 7C-I and 7C-II (triangles) and
7C-III (squares) complete samples in which only the 323 objects deemed
FRIIs are plotted (see Section 2). Filled symbols are quasars or
broad line radio galaxies and open symbols are radio galaxies. We have
binned the $L_{151}-z$ plane in luminosity and redshift, of which 6
bins are well populated by sources from the complete samples. The
percentage of quasars along with associated Poisson errors in each bin
are shown in the bottom-right corner of the plot. The left plot shows
all the FRII sources in the complete samples, whereas the right plot
shows only those in the high emission-line luminosity category (see
text for details). Note that there is one additional radio galaxy in
the 7C-III sample which may be at $z>4$ (Lacy et al. 1999).  }}
\end{figure*}

Our sample consists of three completely--identified samples selected at
(similar) low radio frequencies. These samples have different flux
limits to provide broad coverage of the radio luminosity--redshift
($L_{151}-z$) plane (see Figure \ref{fig:qf1}). This enables the
separation of evolutionary-- and luminosity--dependent effects.

Our faintest sample is the 7C Redshift Survey which is briefly
described here (see also Willott et al. 1998a; Willott et al. in
prep.; Blundell et al. in prep; Lacy et al. 1999). The 7C Redshift
Survey includes all sources with flux-densities at 151 MHz
$S_{151}\geq 0.5$ Jy in three selected regions of sky (7C-I, 7C-II and
7C-III with a total sky area of 0.022 sr). The sample contains 130
radio sources which have all been identified with an optical/near-IR
counterpart. The spectroscopic completeness is $\approx 90$\% with
most of the remaining sources having redshifts well-constrained by
optical/near-IR photometry (Willott, Rawlings \& Blundell, in prep.).

The bright sample used is the 3CRR sample of Laing, Riley \&
Longair (1983), which has complete redshift information for all 173
sources, selected with $S_{178}\geq 10.9$ Jy. 3C 231 (M82) is excluded
because it is a nearby starburst galaxy and not a radio-loud
AGN. 3C 345 and 3C 454.3 are flat-spectrum quasars which are excluded on
the grounds of Doppler-boosting of compact components being responsible 
for raising their fluxes 
above the selection limit. One quasar from the 7C sample
(5C7.230) was excluded for the same reason. 

The intermediate flux-density sample used is a revision of the 6C
sample of Eales (1985) (see Rawlings, Eales \& Lacy in prep.,
hereafter REL, for details). The flux limits of this sample are
$2.0\leq S_{151}<3.93$ Jy and the sky area covered is 0.103 sr. Only 2
of the 58 sources in this sample do not have redshifts determined from
spectroscopy. One object is faint in the near--infrared ($K > 19$), so
we take it as a galaxy at $z=2.0$ in this paper. The other is
relatively bright in the near-IR but very faint in the optical. Its
red colour suggests a galaxy with a redshift in the range $0.8<z<2$
and we assume $z=1.4$ here. Full details of the sample, including
optical spectra, will appear in REL.

All the sources in two of the three 7C regions, 7C-I and 7C-II, and in
the 6C sample have been imaged in the near--infrared $K$-band. This
has enabled the discovery of a few lightly reddened quasars, which
would otherwise have been identified as galaxies on the basis of
optical spectroscopy (see e.g. Willott et al. 1998a). In addition we
have obtained near--infrared spectroscopy of some possible reddened
quasars in these samples to search for broad emission lines (Willott
et al. in prep.). On the basis of the available imaging and
spectroscopy, all sources have been classified into one of three
categories: quasar, broad line radio galaxy (BLRG) or radio
galaxy. Given the relatively small number of reddened quasars found in
the 6C, 7C-I and 7C-II samples, we do not believe the lack of
near--infrared data for the 7C-III sample has caused a
statistically--significant number of quasars to be mis--classified.
Spectra of the quasars and BLRGs in 7C-I and 7C-II are presented with
a detailed discussion of the classification scheme in Willott et al. (1998a).

Our total sample comprises 356 sources (one object, 3C 200, is common
to both the 3CRR and 7C-II samples). We do not exclude compact
symmetric objects (CSOs) from our analysis as previous authors have
(e.g. Barthel 1989, Singal 1996), since we have no a priori reason to
expect them not to be part of the unified schemes. In Section
\ref{pzplots} we repeat our analysis excluding the CSOs to check this
assumption. Sources with FRI radio structure are however excluded.
The main reason for this was a practical one: most (24 out of 33) of
the FRIs are from the 3CRR sample, and the existing spectrophotometry
of these objects is inadequate for the quantitative investigations of
this paper. However, as discussed fully in Section 4.5, the lack of
quasars with FRI radio structure (e.g. Falcke, Gopal-Krishna \&
Biermann 1995), and the concentration of these objects at low
$L_{151}$ and low $z$ (see Fig. \ref{fig:qf1}) makes it easy to
investigate the effects of the exclusion of FRIs on our study of the
quasar fraction.  Although the radio structures of the quasars 5C6.264
and 3C 48 are arguably FRI, we count them in this paper as FRIIs
because of their high radio luminosities. Indeed all objects that
could not unambiguously characterised as FRI or FRII (e.g. the `DA'
sources in the notation of Blundell et al. in prep.)  are also counted
as FRIIs in this study.  Thus our total combined sample contains 323
objects deemed to be FRIIs.

\section{The quasar fraction}
\label{pzplots}

We now investigate the fraction of objects which show broad emission
lines (referred to hereafter simply as the quasar fraction) in the
complete radio samples. Broad line radio galaxies (broad line objects
with $M_B>-23$) are counted as quasars in this analysis, because they
are most likely to be weak quasars viewed within the torus opening
angle (Laing et al. 1994; Hardcastle et al. 1998; Fig. 6 of Willott et
al. 1998a). However, objects with scattered broad lines seen only in
polarised light (4 cases) are counted as galaxies, because the fact
that the broad lines are not observed directly indicates that the
nuclear regions are heavily obscured, most likely by the torus. Since
spectropolarimetry is unavailable for most sources, there is clearly
some danger that the split between weak and scattered-light quasars is
imperfect. A further concern is the question of the classification of
lightly-reddened broad-line objects in which it remains uncertain
whether (a) the reddening is related to the torus or is caused by dust
further from the nucleus, for example in the host galaxy of the
quasar; and (b) whether, particularly given the inhomogeneous
spectrophotometric dataset for the 3CRR sample, all such objects have
yet to be discovered. Taking $0 \ltsimeq A_{V} \ltsimeq 5$ as a
working definition of `light' reddening, we have chosen to count all
such objects known to us as quasars. We will return to this important
point in Section 4.2.

In Fig. \ref{fig:qf1} (left), we plot low-frequency radio luminosity
$L_{151}$ against redshift $z$ for the combined sample of FRII
sources. The $L_{151}-z$ plane has been binned and the percentage of
quasars in each bin shown, along with the associated Poisson errors.
The first thing to note is that there are very few quasars at
$\log_{10} (L_{151} /$ W Hz$^{-1}$ sr$^{-1}) < 26.5$. Excluding this
region, the quasar fraction appears to increase slightly as a function
of both redshift and radio luminosity, ranging from 0.3 to
0.5. However, with Poisson errors of $\sigma \sim 0.08$ these
differences can at best be called marginally--significant, and we will
not consider them further. Note also
that for the high-luminosity bins, the median luminosity in each bin
changes with redshift, so the effects of redshift and luminosity have
not been completely separated.

Laing et al. (1994) proposed that by excluding low--excitation radio
galaxies (LEGs; Hine \& Longair 1979) which have very weak or absent
emission lines and low--ionization narrow lines, the quasar fraction
in 3CRR was not a function of luminosity (or redshift). Their
classification for LEGs was that they have [OIII] to H$\alpha$ ratios
of $<0.2$ and [OIII] equivalent widths of $<3$ \AA. Since the radio
galaxies in the combined samples here span a large range of redshift,
Balmer and/or [OIII] lines are often not observed in optical spectra
and a classification scheme such as this cannot be applied. Instead we
separate sources into those with high and low [OII] emission line
luminosities. The division between these classes we adopt is
$\log_{10} (L_{\rm [OII]} / {\rm W}) = 35.1$. Below this line
luminosity there are very few broad line objects in the complete
samples (Willott et al. 1999).  This cut in narrow emission line
luminosity was chosen to be equivalent to a rest-frame [OII]
equivalent width of 10 \AA~ for a quasar with
$M_{B}=-23$\footnotemark. Note that this division puts more sources in
the low-luminosity category than those typically classified as LEGs.

\footnotetext{The mean rest-frame equivalent width of the [OIII] line
for BQS quasars in the study of Miller et al. (1992) is 30
\AA. Assuming a ratio of 3:1 for the [OIII]:[OII] equivalent widths,
this translates to 10 \AA~ for [OII]. This value was used in Willott
et al. (1999) to determine the strength of the photoionizing continuum
in radio sources. However, the typical ratio between [OII] and [OIII]
may actually be smaller than this, causing Willott et al to have overestimated
the mean [OII] equivalent width by a factor of a few. Baker et al.
(1999) finds a median [OII] equivalent width of 4 \AA~ in the Molongo
Quasar Sample. Also, in the LBQS composite quasar spectrum of Francis
et al. (1991) the [OII] equivalent width is 2 \AA, which would give a
line luminosity of $\log_{10} L_{\rm [OII]} = 35.1$ for a $M_{B}=-25$
quasar.}

On the right-hand side of Fig. \ref{fig:qf1} we plot only sources with
[OII] luminosities $\log_{10} (L_{\rm [OII]} / {\rm W}) \ge 35.1$. The
differences between the quasar fractions in the bins is reduced
somewhat here. Now we find that (with the exception of the
poorly--populated low radio luminosity bin) the quasar fraction is
0.40 in all the bins at the 1$\sigma$ level. The largest change has
occurred in the low-redshift, intermediate luminosity bin which
contains many weak-lined 3CRR galaxies. Note that the $1 \le z < 2$,
intermediate luminosity bin has the lowest quasar fraction. A possible
reason for this is that this bin contains most of the galaxies without
lines in their optical spectra which may not all truly lie within this
redshift range. Considering all 216 high emission line luminosity
sources plotted on the right-hand side of Fig. \ref{fig:qf1}, we find
a quasar fraction of $0.40 \pm 0.03$ implying a mean torus
half-opening angle of $53^{\circ}\pm 3 ^{\circ}$ for the luminous
population. There is likely to be quite a range of torus opening
angles and the small error presented here on the mean angle should not
be interpreted as the dispersion in opening angles present in the
population.

We have repeated the analysis of this section excluding CSO sources
(projected linear sizes $\leq 30$ kpc). We find that this reduces the
quasar fraction by $\approx 0.04$ in all the bins. The quasar fraction
of all luminous CSO sources is $0.56 \pm 0.08$, different at the
2$\sigma$ level from that of non-CSO sources. We defer discussion of possible
reasons for a higher quasar fraction for CSOs in low-frequency selected
samples to a future paper.

\section{Possible causes of the change in quasar fraction with luminosity}

\subsection{Selection effects caused by lower intrinsic broad line luminosity}

We first investigate whether the decrease in quasar fraction
with decreasing narrow-line and radio luminosity is a simple
selection effect. If the low-luminosity objects have lower luminosity
narrow emission lines it is natural to expect that their 
broad line and continuum luminosities should be 
lower too (e.g. Miller et al. 1992).
It follows that their spectra may on average have lower
signal-to-noise and weak broad lines may be missed. Given the
strong correlation between narrow-line and radio luminosities
(e.g. Willott et al. 1999) this could
explain the deficit of broad line objects at radio luminosities
$\log_{10} (L_{151}$ / WHz$^{-1}$sr$^{-1}) <26.5$, and also the
apparent constancy of the quasar fraction for objects with luminous
narrow lines discussed in Section 3.

To test whether selection effects such as these could cause the
apparent lack of low-luminosity broad line objects, we first calculate
the expected broad line fluxes of objects in the 3CRR and 7C samples
as a function of luminosity. To do this we make use of the fact that
the narrow emission line luminosities of radio galaxies and quasars
are positively correlated with low-frequency radio luminosity as shown
in Willott et al. (1999). This correlation has a slope of $0.79 \pm
0.04$ in the sense that $L_{\rm [OII]} \propto L_{151}^{0.79}$ and is
most likely due to a correlation between jet power and photoionizing
continuum luminosity (Rawlings \& Saunders 1991). Use of this relation
enables one to estimate the narrow line [OII] luminosity for values of
radio luminosity.

A similar correlation between the broad line luminosity and radio
luminosity is expected to hold. There are several lines of evidence
supporting this. First, radio luminosity is correlated with the
optical continuum luminosity for steep-spectrum quasars (Serjeant et
al. 1998) and quasars show little variation in broad line equivalent
widths over a wide range of luminosity (e.g. Osmer \& Shields
1999). Second, Celotti et al. (1997) have shown that broad line
luminosities are correlated with the power in pc-scale jets in
radio-loud quasars.

To determine the relationship between the strengths of narrow and
broad lines in quasar/BLRG spectra, we adopt the line ratios from
the composite quasar spectrum of Francis et al. (1991). Although this
is from an optically-selected sample, we do not expect significant
differences between the line ratios of radio-loud and radio-quiet
quasars. This is borne out by a comparison with the Molonglo Quasar
Sample of Baker \& Hunstead (1995): repeating the analysis using their
line ratios gives a virtually identical result.

\begin{table}
\footnotesize
\begin{center}
\begin{tabular}{cllcc}
\hline\hline
\mc{1}{c}{$\log_{10} L_{151}$}   & \mc{1}{l}{$z_{\rm 3CRR}$}& \mc{1}{l}{$z_{\rm 7C}$}& \mc{1}{c}{broad line}& \mc{1}{c}{broad line} \\
\mc{1}{c}{(WHz$^{-1}$sr$^{-1}$)} & \mc{1}{c}{}        & \mc{1}{c}{}        & \mc{1}{c}{(3CRR)} &  \mc{1}{c}{(7C)}  \\
\hline\hline  
28.0 & 1.2  & 4.0  & MgII      & Ly$\alpha$ \\
27.5 & 0.60 & 2.6  & MgII      & Ly$\alpha$ \\
27.0 & 0.35 & 1.8  & H$\beta$  & CIV        \\ 
26.5 & 0.20 & 1.0  & H$\alpha$ & MgII       \\
26.0 & 0.13 & 0.60 & H$\alpha$ & MgII       \\
25.5 & 0.06 & 0.35 & H$\alpha$ & H$\beta$   \\
25.0 & 0.04 & 0.20 & H$\alpha$ & H$\alpha$  \\

\hline\hline  
\end{tabular}
\end{center}              
{\caption[Table of observations]{\label{tab:bl3c7c} This table shows
typical redshifts of sources from the 3CRR and 7C samples (columns 2
and 3) at various radio luminosities (column 1). Columns 4 and 5 show
the strongest broad lines observable over the relevant optical
wavelength ranges corresponding to these redshifts. }} \normalsize
\end{table}

We consider a range of radio luminosities and attach to each one
typical redshifts at which sources of these luminosities are observed
in both the 3CRR and 7C samples. The correlation between the [OII]
line and radio luminosity is then used to determine typical [OII]
luminosities at each value of radio luminosity. From the typical
wavelength range covered by optical spectra (4000\AA -- 8500\AA), we
find the strongest broad line observable at each redshift.  The
typical luminosities, redshifts and strongest broad lines are shown in
Table \ref{tab:bl3c7c}. Finally, we use the ratio of broad to [OII]
lines in Francis et al. (1991) to calculate the luminosity and hence
flux of the strongest broad line expected at each luminosity-redshift
pair.

There is considerable scatter in the relationship between narrow
emission line luminosity and radio luminosity ($\sigma=0.5$
dex). Extra scatter comes in due to the conversion from narrow to
broad line luminosities. Therefore a total scatter of $1\sigma=0.6$
dex was assumed to determine the fraction of objects with broad line
fluxes significantly below the characteristic values. 

\begin{figure}
\hspace{-0.25cm} \epsfxsize=0.48\textwidth \epsfbox{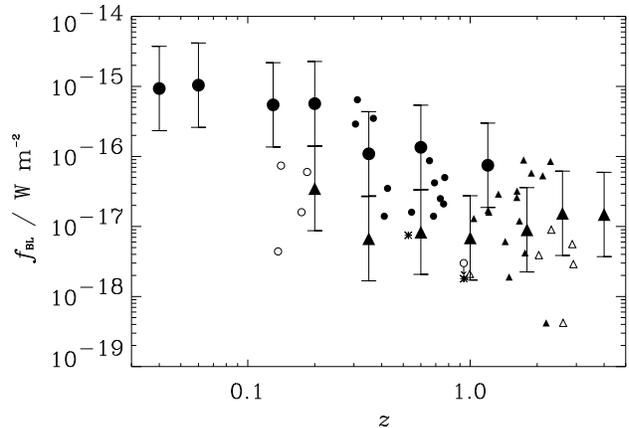}
{\caption[junk]{\label{fig:lineflx} Expected broad line fluxes for
`typical' 3CRR (large circles) and 7C (large triangles) sources as a
function of redshift according to the radio-optical correlation as
described in Section 4.1. The error bars show approximate $1\sigma$
deviations (i.e. $\approx$ 1 in 6 sources should fall below the bottom
of the error bars). The small symbols show the actual data for 3CRR
(circles) and 7C (triangles) quasars and the 7C BLRGs
(asterisks). Open triangles are 7C quasars which are known to be
reddened. The open circles show the reddened quasar 3C 22 at $z=0.93$
(broad MgII upper limit from Rawlings et al. 1995) and the 3C BLRGs
from Hill et al. 1996 (see Section 4.2).}}
\end{figure}

Figure \ref{fig:lineflx} shows the expected flux of the strongest
broad line observable at each redshift as listed in Table
\ref{tab:bl3c7c} for both 3CRR and 7C. Also plotted are the actual
data for the 7C-I and 7C-II quasars and BLRGs (Willott et al. 1998a)
and all 3CRR quasars with $z<0.86$ in the RA range 18--12 hr (all from
Jackson \& Browne 1991 except 3C 109 from Goodrich \& Cohen 1992). In
general points corresponding to 3CRR BLRGs are not plotted because
only a few of their broad line fluxes are available in the literature,
although a few points from Hill et al. (1996) are plotted to
illustrate the discussion of Section 4.2. The quasars which are
red(dened) [$\alpha_{\rm opt}>1$, where $f_{\nu} \propto
\nu^{-\alpha_{\mathrm opt}}$] are shown as open symbols on
Fig. \ref{fig:lineflx}; again the 3CRR spectrophotometric dataset is
too inhomogeneous for this to be done reliably for 3CRR sources. Note
that very few quasars fall below the limits of the model error
bars. Virtually all of those which do are reddened (the exception is
5C6.282 at $z\approx2$, which has an unusual spectrum with very narrow
Ly$\alpha$; see Willott et al. 1998a). The similarity of the predicted
fluxes with those observed justifies the use of the line ratios and
method described here.

The key point to note from Fig. \ref{fig:lineflx} is that as one
considers lower radio luminosities in each sample (i.e. lower
redshifts), the expected broad line fluxes {\em increase}. Therefore,
so long as the spectra of lower redshift sources have similar
exposure times to those at higher redshift, the broad lines should
be visible if they are there and unreddened. 
For the 3CRR sample the spectra are
rather inhomogeneous and it is difficult to prove that this is the
case (but note that the deficit of 3CRR sources is most marked at
$z<0.3$, where the H$\alpha$ line is covered by optical spectra and
the expected broad H$\alpha$ fluxes are high). However, in the 7C
sample, similar exposure times have been attained for the spectra
at low redshift to those at high redshift. Therefore we are confident
that in the absence of significant reddening of the BLR, an
insignificant number of sources would be mis-classified.

\begin{figure}
\hspace{-0.25cm} 
\epsfxsize=0.48 
\textwidth 
\epsfbox{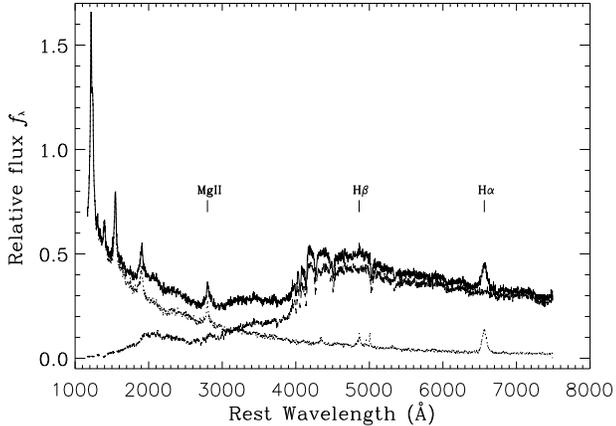}
{\caption[junk]{\label{fig:onespec} Superposition of a blue 
model quasar/BLRG spectrum and a red galaxy spectrum 
as described in the text. The quasar is two 
magnitudes fainter in $B-$band than the galaxy. The upper line shows the sum of
the two components. Note that the broad MgII and H$\alpha$ lines are
clearly visible in the total flux spectrum, the former because of the
quasar dominance at wavelengths $\lambda < 3000$ \AA~ and the latter because of
its high equivalent width.}}
\end{figure}

A further problem is whether a weak quasar spectrum can be
discerned against a stronger host galaxy spectrum. To determine if
this could cause broad line selection problems, synthetic spectra of
quasars and galaxies were created and combined. The model used for the
galaxy spectra was a 1 Gyr old stellar population synthesis model of
Bruzual \& Charlot (1993). For the quasar spectra, the LBQS composite
of Francis et al. (1991) was used. Poisson noise was added and
it was found that broad H$\alpha$ or MgII should still be clearly
visible for a quasar $2$ magnitudes (in $B$-band) fainter than the
galaxy (Fig. \ref{fig:onespec}). Broad H$\beta$, however, is more
difficult to detect.  Spectra with poor blue wavelength coverage
(e.g. no data below $5000$ \AA) of sources at $0.3<z<0.8$ would
not include H$\alpha$ or MgII, in which case H$\beta$ is the brightest
line. Almost all the 7C and 6C spectra cover a large enough wavelength
range to avoid this problem. However many spectra of 3CRR sources in
the literature have a smaller wavelength range. But the radio--optical
correlation (and Table \ref{tab:bl3c7c}) show that quasars at these
redshifts in 3CRR typically have high luminosities and should not be
fainter (at $B$-band) than their host galaxies. Indeed, the deficit of
broad line objects in 3CRR only really begins at $z \ltsimeq
0.3$. Hence quasar broad lines should be clearly visible against
typical background galaxy spectra down to $\log_{10} (L_{151}$ /
WHz$^{-1}$sr$^{-1}) \approx 25.5$, with possible problems only at
fainter radio luminosities.

Another possibility to consider is that the narrow lines may be much
stronger than the broad lines, so only weak wings in the line profiles
would be seen. However, this picture seems unlikely for many sources
in the absence of broad line reddening, since if both the narrow and
broad lines are photoionizied by the same nuclear source it would
require a conspiracy between broad and narrow line covering factors.

We conclude that the low quasar fraction at low luminosity
is not due to selection effects caused by the difficulty of 
detecting intrinsically less luminous broad lines.

\subsection{Increased reddening at low luminosities}

In this section we consider the possibility that the drop in quasar
fraction at low luminosities is due, at least in part, to an increase
in the fraction of lightly-reddened ($0 \ltsimeq A_{V} \ltsimeq 5$)
lines-of-sight as quasar luminosity decreases.  It can be seen from
Fig. \ref{fig:lineflx} that the majority of the 7C quasars are lightly
reddened at the highest redshifts. We exclude from this paper any
discussion of any systematic changes of quasar reddening with cosmic
epoch.

The question of the range of reddening towards the nuclei of powerful
(3CRR) $z \sim 1$ radio sources has recently been addressed by the
$3.5\mu$m imaging programmes of Simpson, Rawlings \& Lacy (1999) and
Simpson \& Rawlings (2000). They concluded that these high luminosity
nuclei are mostly either naked ($A_{V} \sim 0$) or heavily obscured
($A_{V} \gtsimeq 15$) with only a small fraction ($\sim 15$ per cent)
of intermediate lightly-reddened cases. If these results could be
extrapolated to the lower narrow-line and radio luminosity regimes
then light reddening of quasar nuclei would not be a serious issue for
our study of the quasar fraction.

However, two studies suggest that the fraction of lightly-reddened
lines-of-sight might increase significantly at lower narrow-line and
radio luminosities. First, Hill, Goodrich \& DePoy (1996) studied the
Pa$\alpha$ and $H \alpha$ lines in a complete sample of low redshift
($0.1 <z< 0.2$) 3CR sources, looking for evidence of reddened broad
lines in radio galaxies. These objects lie near the top of
lowest-$L_{151}$ (and $z$) bin in Fig. \ref{fig:qf1}.  They found
broad lines in six objects out of a complete sample of thirteen
(including two FRI sources), of which three had intermediate values of
$A_{V}$. This study, although subject to a major problem with small
number statistics together with the difficulties of detecting lightly
reddened broad lines over certain ranges of redshift (see
Fig. \ref{fig:lineflx}), hints that an increase in the fraction of the
lightly-reddened population at low luminosities might provide part of
the explanation for the dropping quasar fraction at low
luminosities. Second, Baker (1997) showed that a typical
low-luminosity lobe-dominated quasar, i.e. an object akin to those
near the top of lowest-$L_{151}$ (and $z$) bin in Fig. \ref{fig:qf1},
has a reddening $A_{V} \sim 3$ and is thus lightly reddened.

We conclude that some of the drop in the quasar fraction at low
luminosities might be explained by an increased chance of light
reddening towards low luminosity quasar nuclei. In other words, we
cannot allay the suspicion that some lightly reddened quasars have
been misclassified as galaxies on the basis of existing, chiefly
optical, spectroscopy. In the lowest-$L_{151}$ and $z$ bin of
Fig. \ref{fig:qf1} this is a problem for (very low-$z$) 3CRR sources
because of the inhomogenous optical dataset; it is a problem for the
6C and 7C sources because their redshifts are sufficient to remove the
strong H$\alpha$ line from the optical window (see
Fig. \ref{fig:lineflx}). To pursue this idea further we next consider
specific physical models which might explain this behaviour.

\subsection{The receding torus model}

Lawrence (1991) proposed that the apparent increase in quasar fraction
and hence $\theta_{\rm trans}$ with luminosity in the 3CRR sample
could be explained naturally by a `receding torus' model. Dust
sublimates at $T\sim 1500$ K and therefore the inner radius of a dusty
torus/disc may be governed by the radius at which this temperature is
achieved via radiation from the AGN. The more luminous the AGN the
larger this radius and {\em if} the scale height of the torus $h$ is
independent of luminosity, then more luminous sources will have a
larger $\theta_{\rm trans}$; the inner radius of the torus $r$ scales
as $L^{0.5}$ and $\theta_{\rm trans} = \tan^{-1} (r / h)$.  This model
can also explain more lightly-reddened quasars at low luminosity since
a larger fraction of the lines-of-sight pass through intermediate
columns of obscuring material when the inner wall of the torus lies
closer to the nucleus (see e.g. Fig. 8 of Hill et al. 1996).

\begin{figure}
\hspace{-0.25cm} \epsfxsize=0.48\textwidth \epsfbox{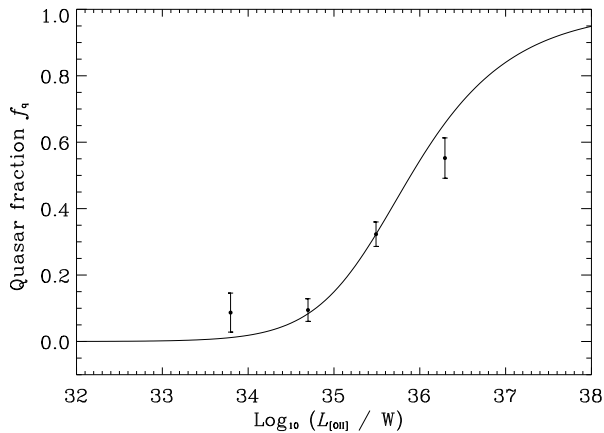}
{\caption[junk]{\label{fig:qf2} The points with Poisson error bars
show the quasar fractions in four bins of [OII] emission line luminosity
plotted against the median $\log_{10} (L_{\rm
[OII]})$ of the sources in each bin. The data used were for the
objects deemed FRIIs in the combined sample 3CRR, 6C and 7C dataset
(see Willott et al. 1999). 
The line is a simple prediction made on the basis of
the receding torus model, normalised at
$\log_{10} (L_{\rm [OII]} / {\rm W})=35.5$, as
described in Section 4.3.}}
\end{figure}

Simpson (1998) has shown that if the
height of the torus is independent of the ionizing luminosity $L$ 
then the receding torus model predicts a
theoretical quasar fraction $f_{\rm q}$ given by
\begin{equation}
\label{eqn:qf}
f_{\rm q} = 1- \left( 1+ \frac{L}{L_{0}} \tan ^{2} \theta_{0} \right)
^{-0.5},
\end{equation}
where ${L_{0}}$ and $\theta_{0}$ are the normalisation luminosity and the
torus half-opening angle at that luminosity, both of which are fixed by the
measurement of the quasar fraction at some normalisation luminosity.
We further assume that the [OII] luminosity scales with the ionizing
continuum luminosity, and that it is not significantly affected by the changing
torus opening angle, so that we can replace ionising luminosities in Eqn 
\ref{eqn:qf} by values of $L_{\rm [OII]}$. \footnotemark
\footnotetext{
As emphasised by Simpson, [OII] is probably a far from ideal
choice of narrow emission line for this 
procedure because for 
an ionisation-bounded emission line cloud, the [OII] emission
is from a region whose size depends much less strongly on
the incident photoionising flux than the sizes of the regions responsible for 
higher-ionization lines like [OIII]. This leads to a much weaker dependence
of [OII] luminosity on incident photoionising luminosity $L$ than the 
near proportionality between, say, [OIII] luminosity and $L$.
Unfortunately, the choice of [OII] was
forced on us by the available spectrophotometric
data (see Willott et al. 1999).
}
Binning all FRII sources in terms of $L_{\rm [OII]}$ we
find that for $35<\log_{10} (L_{\rm [OII]} / {\rm W})<36$, the quasar
fraction is $0.32 \pm 0.04$ giving $\theta_{0} =46^{\circ}$. This is
used to normalise Equation \ref{eqn:qf} which is plotted in
Fig. \ref{fig:qf2} along with the
quasar fraction measured in four bins of $\log_{10} (L_{\rm [OII]})$.
 
This simple model appears to fit the data fairly well although it does
predict a marginally sharper transition between low and high quasar
fractions than the data (at about the $1 \sigma$ level). Note that the
above model depends crucially upon the height of the torus being
independent of luminosity. If the height of the torus increases with
luminosity, albeit more slowly than the inner radius, then this would
flatten the model distribution to something more closely resembling
the data. 

Taken together with the results of Hill et al. (1996), we conclude that
the receding torus model provides a plausible  if non-unique explanation
for the declining quasar fraction below 
$\log_{10}(L_{\mathrm [OII]}/{\mathrm W}) = 35.1$.

\subsection{Dust--ejection by jets or outflow}

Tadhunter et al. (1999) have recently obtained high-resolution
near-infrared imaging of the powerful radio
galaxy Cygnus A which shows an edge-brightened biconical structure
roughly aligned with the jets. They suggest that at least some of the
very high reddening towards the nucleus of Cygnus A ($A_V \approx 70$ mag) 
occurs in a kiloparsec scale dust-lane. For radio-loud sources, the jets may
clear a path through this dust, so that only when viewed within
a critical angle of the jet-axis is there a chance of a line-of-sight having
little reddening. The jet power could control
the dust-clearing efficiency in such a way as to give 
an anti-correlation between reddening and radio luminosity.
The conceptual difference between this model and 
the receding torus model is that 
it is now a decrease in jet power (and hence the radio luminosity) rather than
a decrease in optical luminosity that controls the closing down of the
effective opening angle of the radiation cone at low luminosities.
Tadhunter et al. speculate that lightly reddened quasars
are objects viewed within the nuclear torus opening angle, but not
within the region cleared of dust by the jets. Another possibility
(e.g. Dopita et al. 1998) is that dust is cleared by a 
radiation-driven outflow in which the opening angle and strength of 
the outflow could be related to the luminosity of the nucleus and hence
give rise to the observed dependence of quasar fraction on luminosity.

\subsection{Dual-population models}

The evidence of increased reddening at low luminosities (Section 4.2)
and the quantitative success of the receding torus model (Section 4.3)
hint that we have a reasonable explanation for at least part of the
drop in quasar fraction at low luminosities. In this section 
we discuss separate evidence for a second conceptually different mechanism.

A corollary of Fig. \ref{fig:qf2} is that at low luminosities the low
fraction of quasars results from a small value for the typical opening
angle of the torus $\theta_{\rm trans}$.  Falcke et al. (1995) have
cited this as a strength of the receding torus model since if the
opening angle approaches the beaming angle of the jet then there will
be an unobscured view of the nucleus of low power jets only when their
emission is also strongly Doppler-boosted.  Such objects would be
classified as BL Lac objects rather than quasars, providing a neat
explanation for the well-known lack of quasars with FRI radio
structures. We are unaware of any systematic measurements of broad
line fluxes in BL Lac type objects which would provide a quantitative
test of this hypothesis. However, recent observations of BL Lacertae
itself (Corbett et al. 2000), an object with an extended radio flux
characteristic of a low-luminosity radio galaxy (Antonucci \& Ulvestad
1985), suggest that in at least this object there are broad emission
lines at about the predicted level, and that these lines are excited
by a hot accretion disc.

However, other recent results suggest that the small opening angles
required at low luminosities from Fig. \ref{fig:qf2} are more likely
to argue against the generality of the receding torus model.  If it
were true that low luminosity radio galaxies possessed very thick
obscuring tori then it should prove very difficult to get a clear view
of their central regions at optical wavelengths.  Chiaberge, Capetti
\& Celotti (1999) have studied the nuclear emission from FRI galaxies
using the HST. In most cases they find probable optical synchrotron
from the nuclear regions, which they interpret as implying a clear
view of the inner jet and hence a lack of obscuring material on scales
greater than $\sim100$ pc. This straightforwardly rules out a kpc-scale
obscuration model for FRIs along the lines of the model proposed by
Tadhunter et al. (1999) for Cyg A. Still stronger constraints come
from the nearby FRI radio galaxy M87 since its optical synchrotron
core is smaller than $\sim 5$ pc and varies on the timescales of
months (Tsvetanov, Kriss \& Ford 1997), implying that the emission
region is a pc or smaller in size. Since the inner regions of the
optical synchrotron-emitting jet are in clear view, and these size
scales are comparable in size to the expected size of the obscuring
torus in low-luminosity quasars (e.g. Netzer \& Laor 1993), these
results begin to question the existence of any thick torus, at least
in the case of M87. Chiaberge et al. maker stronger statements than
this by making the additional assumption that this optical emission is
emitted from within the region probed by radio VLBI which is argued to
be within $\sim 0.1$ pc of the central black hole (Junor, Biretta \&
Livio 1999). We remain to be convinced that these observations rule
out the existence of a pc-scaled torus in M87.  There is a plethora of
evidence from studies of M87 using optical continuum (Chiaberge et
al. 1999), emission-line (Ford et al. 1994) and hard X-ray (Guainazzi
\& Molendi 1999) observations that it contains a supermassive black
hole accreting at a tiny ($< 10^{-4}$) fraction of its
Eddington-limited rate (see Willott et al. 1999 for a discussion of
this object in the context of the radio-optical correlation).

As shown in Fig. \ref{fig:qf1}, excluding objects with weak narrow
emission lines gives a quasar fraction which is virtually independent
of redshift and/or radio luminosity. This observation can be naturally
explained if the weak-line objects belong to a different population
from the strong emission line objects. Perhaps, as is seemingly the
case for M87, a large fraction of these objects are fundamentally
different in that they lack a well-fed quasar nucleus and also a thick
obscuring torus (Chiaberge et al. 1999), and perhaps their jets are
powered by a fundamentally different mechanism (e.g. Blandford \&
Begelman 1999).  The continuity of the radio-optical correlation for
radio sources provides a weak argument against a fundamental change in
the accretion process (Rawlings \& Saunders 1991), but evidence for
systematic changes in the slope of this relation at low luminosities
(Zirbel \& Baum 1995) may be evidence that there are at least some
differences at low radio luminosities.  Arguing along similar lines,
Laing et al. (1994) proposed that the low-excitation radio galaxies
(LEGs) in their 3CRR sample would lack broad lines if viewed from any
orientation. Recall that we specifically excluded FRI radio sources
from our analysis of the quasar fraction in Section 3. The only
difference that including the FRIs would have made, would be to
decrease the quasar fraction in the lowest luminosity bin from $0.14
\pm 0.04$ to $0.10 \pm 0.03$.

Such a `two population' model would remove another troubling problem
for any extrapolation of the receding torus model to the lowest quasar
luminosities. Seyfert galaxies -- radio-quiet AGN with similar optical
narrow-line luminosities to radio galaxies in the lowest-$L_{151}$ bin
of Fig. \ref{fig:qf2} -- do not have the narrow opening angles
predicted by the receding torus model.  Direct observations of the
ionization cones in nearby Seyfert galaxies typically show cones with
half-opening angles in the range $20-50^{\circ}$ (Wilson \& Tsvetanov
1994), considerably greater than those predicted. Moreover, the vast
majority of Seyfert galaxies do not seem to be obscured, i.e. of type
2, as would be required by narrow opening angles. Hence, as noted by
Falcke et al (1995) it would appear that the receding torus model
cannot apply similarly to radio-quiet and radio-loud AGN requiring
some ad-hoc explanation, for example a different geometry for the
obscuring material (i.e. a smaller scale height), due to a difference
in black hole mass, environment and/or angular momentum.

We can make a crude association of the putative second population of
radio sources with all sources with narrow emission line luminosities
$\log_{10} (L_{\rm [OII]} / {\rm W}) < 35.1$. Due to the scatter in
the radio--optical correlation, it is not expected that this would
lead to a clean division between the two populations at a particular
[OII] luminosity, and if the data were available, a classification
based on line ratios, and hence excitation, might well prove
cleaner. However, subject to this limitation, the change in quasar
fraction with redshift of a combination of the two populations (seen
on the left plot of Fig. \ref{fig:qf1}) is then naturally explained by
the less rapid cosmic evolution of the low-luminosity population
(e.g. Urry \& Padovani 1995) than the high-luminosity population
(shown on the right plot of Fig. \ref{fig:qf1}). Hence at
high-redshift, the contribution of the low-luminosity population is
negligible and the quasar fraction of 0.4 is just that of the
high-luminosity population. Dual population modelling of the
low-frequency radio luminosity function is consistent with just such a
scheme (Jackson \& Wall 1999; Willott et al. in prep.). Also, Laing et
al. (1994) and Hardcastle et al. (1998) find that low-excitation radio
galaxies have linear size and core prominence distributions consistent
with an isotropic population.

There are two residual concerns with the two population model. First,
many of the low-luminosity objects have emission lines, so their
excitation and the correlation between luminosity and ionization
parameter (Saunders et al. 1989; Tadhunter et al. 1998) need some
explanation. Second, it must explain why radio galaxies and quasars
have different narrow line luminosity distributions at intermediate
luminosities (Jackson \& Browne 1990) but similar distributions at
high luminosities (Jackson \& Rawlings 1997). The first concern can be
addressed by simple analogy with M87: the absence of a broad-line
quasar nucleus does not mean an absence of a photoionising source; for
example Dopita et al. (1997) favour radiative shocks as a source of
the excitation for the H$\alpha$ lines in the nuclear disc of M87, and
shock models in which ionization parameter correlates with line
luminosity are easily envisaged (e.g. Dopita \& Sutherland 1995).  The
second concern is probably also easily dealt with.  Considering first
the 3CRR sample, the dual-population model is consistent with the drop
in quasar fraction at low luminosities because a low luminosity 3CRR
source is necessarily at low redshift where there is clearly a mixture
of both populations; at high redshifts only the high-luminosity
population is observed.  Thus any comparative narrow emission line
study of quasars and radio galaxies which is based on bright radio
samples (Jackson \& Browne 1990; Jackson \& Rawlings 1997) should
yield different line luminosity distributions at low redshift, and
similar distributions at high redshift, which is just as observed.
However, a small difference in the distributions of the [OII] line
luminosities of intermediate luminosity quasars and radio galaxies is
also observed in the 7C sample (Willott et al. 1999). The lower radio
flux limit of this sample means that these intermediate luminosity
sources are at high redshift, where there are few low emission line
luminosity objects, so the difference is not due to the mixing of the
two populations.  This is most likely the result of small but
inevitable biases pointed out by Rawlings \& Saunders (1991), and
quantified in the context of the receding torus model by Simpson
(1998): a positive correlation between quasar luminosity and opening
angle, coupled with inevitable scatter in quasar luminosity at a fixed
radio luminosity, means that objects viewed within the opening angle
are biased towards the more luminous objects within the scatter.

\subsection{Time variability}

There is yet one more possible cause of the drop in quasar fraction
at low luminosities: systematic differences in the time
variability of objects with luminosity.

In Willott et al. (1999) we used the narrow emission line--radio
correlation to suggest that the most luminous objects are probably
accreting at rates close to the Eddington limit, but lower luminosity
sources are sub-Eddington accreters.  It therefore seems plausible
that the lower luminosity sources have more scope for variability,
since an object accreting at the Eddington rate should have a ready
fuel supply which is accreted at a fairly steady rate limited by
radiation pressure. In contrast, sub-Eddington accretion suggests
there is not a ready supply of fuel available and hence fluctuations
in accretion rate may be more likely. Indeed observations of
radio-quiet quasars show that the lower luminosity quasars are more
highly optically-variable than higher luminosity quasars over
timescales of a few years (e.g. V\'eron \& Hawkins 1995; Cristiani et
al. 1996; Paltani \& Courvoisier 1997), although it should be noted
that some authors attribute at least some of this variability to
gravitational microlensing (e.g. Hawkins \& Taylor 1997).

The small quasar fraction at low luminosities could be explained by
these objects spending a significant fraction of their active
lifetimes in a `quiet' state whereas the more luminous objects are
continuously active over their entire lifetime.  Due to light travel
time effects, different emission regions of quasars have different
variability timescales: BLR $\sim$ months; NLR $\sim 10^{4}$ yr; radio
lobes $\sim 10^{6}$ yr. Reverberation mapping of quasars has shown
that the broad line fluxes follow the nuclear continuum flux with just
such a time lag (see Peterson 1993 for a review). If a quasar
undergoes high-amplitude variability over a timescale $\sim 100$ yr,
then only the continuum and BLR fluxes would be observed to undergo
this strong variability, and the NLR and extended radio emission would
simply reflect the time-averaged output of the central engine. Note
that a quasar undergoing a {\em decrease} in luminosity of this sort
of timescale may then appear as a low-excitation radio galaxy. If this
behaviour occurs only in a significant fraction of the intermediate
and low luminosity objects, then it could reproduce the observed
change in quasar fraction.

Essentially this model is a slightly more complicated version of the
two-population model considered in Section 4.5, with the two
populations being unified along a time rather than a jet orientation
axis. It is important to stress that there is as yet no firm evidence
for the type of time variability required.  However, the few long-term
time variability studies hint that there may be some interesting
effects: the BLRG 3C 390.3 was found to undergo a sustained decrease
of 1.5 mags in the optical over $\approx 80$ years (Cannon, Penston \&
Penston 1968); however, Angione (1973) used archival plates to study
the variability of 23 quasars over $\approx 60$ years finding no
evidence for long-term sustained increases or decreases.

A separate issue related to time variability is whether the
probability of individual radio sources being observed to be quasars
varies systematically throughout their lifetimes. Because flux-limited
samples introduce inevitable biases into the age distributions of the
sources they contain (e.g. Blundell, Rawlings \& Willott 1999), this
can lead to subtle effects.

\section{Conclusions}

Using complete, low-frequency selected samples of radio-loud AGN we
have investigated the fraction of observed broad line objects --
the quasar fraction -- as a function of
redshift, and radio and emission line luminosity. Our findings are
interpreted in terms of orientation-based unified schemes. We find that

\begin{itemize}
\item{Considering only those sources with
strong narrow emission lines [$\log_{10}
(L_{\rm [OII]} / {\rm W}) \ge 35.1$, corresponding roughly to
$\log_{10}(L_{\mathrm 151}/{\mathrm W}) \ge 26.5$], 
the quasar fraction is virtually
independent of radio luminosity and redshift, and is consistent with a
simple unified scheme involving an obscuring torus with a
half-opening angle $\theta_{\rm trans}$ of $53^{\circ}$.}
\item{For less luminous emission-line sources, 
the quasar fraction is much lower, a finding which is not 
simply the result of selection effects induced by the 
lower intrinsic broad line luminosities of these sources. 
}
\end{itemize}

We have found evidence which supports at least two probable
physical causes for the drop in quasar fraction at low luminosity:
(i) a gradual decrease in $\theta_{\rm trans}$ and/or a gradual 
increase in the fraction of lightly-reddened ($0 \ltsimeq A_{V} \ltsimeq 5$) 
lines-of-sight with decreasing quasar luminosity; and (ii)
the emergence of a second population of low luminosity radio sources 
which, like M87, lack a well-fed quasar nucleus, and may well
lack a thick obscuring torus.

\section*{Acknowledgements}

We would like to thank Steve Eales, Gary Hill, Julia Riley and David
Rossitter for important contributions to the 7C Redshift
Survey. Thanks also to Robert Laing and Chris Simpson for some very
useful discussions, and to the referee Ian Browne for useful
suggestions.  This research has made use of the NASA/IPAC
Extra-galactic Database, which is operated by the Jet Propulsion
Laboratory, Caltech, under contract with the National Aeronautics and
Space Administration. CJW thanks PPARC for support.

\end{document}